\documentclass[aps,prd,twocolumn,superscriptaddress]{revtex4-2}
\pdfoutput=1

\usepackage{amsmath,amssymb,slashed}
\usepackage{graphicx}
\usepackage{xcolor}
\usepackage[colorlinks=true,citecolor=blue,linkcolor=blue,urlcolor=blue]{hyperref}

\begin{document}

\title{Right-Handed Leptonic Mixing and Enhancement Band in Left-Right Symmetry}

\author{Vladimir Tello}
\affiliation{University of Split, FESB, Croatia}
\date{\today}

\begin{abstract}
   
   Left--right (LR) symmetric theories predict right-handed charged currents whose flavor structure encodes the realization of parity. While the right-handed quark mixing matrix closely tracks its left-handed counterpart, the leptonic sector with purely Dirac neutrinos has remained structurally unclear.
We show that, in contrast to the quark case, parity in the Dirac leptonic sector admits a localized, branch-dependent enhancement band in which RH--LH misalignment becomes parametrically large despite small parity breaking. We derive analytic solutions of the LR consistency equation and demonstrate that the interplay between spontaneous parity violation and spectral near-degeneracies leads to a qualitatively new pattern of right-handed mixing.
This establishes the Dirac leptonic sector of the minimal LR model as a predictive and structurally distinct regime.

\end{abstract}

\maketitle

\section{Introduction}

Left--right (LR) symmetric theories provide a well-motivated extension of the Standard Model,
restoring parity at high energies and relating left- and right-handed charged currents
\cite{Pati:1974yy,Mohapatra:1974gc,Senjanovic:1975rk,Senjanovic:1978ev}.
In this framework, the flavor structure of right-handed (RH) interactions is in principle  constrained by the underlying symmetry.

In the quark sector, this relation has been understood in detail: despite sizable parity breaking,
the RH mixing matrix closely tracks its left-handed counterpart, a consequence of the strong
hierarchy of quark masses \cite{Senjanovic:2014pva,Senjanovic:2015yea}.
A natural open question is whether an analogous behavior holds in the leptonic sector.

The answer depends crucially on the nature of neutrino masses. 
While LR symmetry naturally accommodates Majorana neutrinos through seesaw
realizations \cite{Minkowski:1977sc,Yanagida:1979as,Mohapatra:1979ia,Glashow:1979nm,GellMann:1980vs},
Dirac neutrinos are equally consistent from the viewpoint of the symmetry,
in particular in the minimal doublet-breaking realization \cite{Branco:1978bz}.
In this case, however, the leptonic sector has received comparatively less attention
as a predictive and self-contained framework.

In this work we adopt a minimal viewpoint and take the Dirac LR model at face value,
without addressing the origin of the small neutrino masses.
Instead, we investigate the structural consequences implied by parity symmetry.

We show that, in contrast to the quark case, the leptonic sector exhibits a qualitatively
different behavior. The LR consistency conditions reduce to a nonlinear relation whose
solutions organize into discrete branches. In mixed-sign branches, the interplay between
small parity breaking and small neutrino mass splittings leads to a parametric enhancement
of RH--LH mixing differences. As a result, sizable misalignment can arise despite the
smallness of the symmetry-breaking parameter $\epsilon$, while outside this localized
band the mixing remains perturbatively aligned.

In this framework, the RH leptonic mixing matrix is fixed
in terms of the left-handed one, the neutrino spectrum, and a single spontaneous
parity-breaking parameter $\epsilon$, up to discrete sign assignments.

We analyze this structure analytically in complementary regimes and solve the exact
matrix equation numerically in the $(m_{\nu,\text{lightest}},\epsilon)$ plane,
providing a unified description of the enhancement mechanism across parameter space.

It is useful to clarify the relation of the present analysis to earlier work.
The parity-based reconstruction of RH mixing, including the quark-sector treatment
and the small-$\epsilon$ expansion, was developed in
Refs.~\cite{Senjanovic:2014pva,Senjanovic:2015yea};  this provides the direct
methodological starting point for the present analysis. Related LR-symmetry constraints on leptonic Dirac mass matrices were studied in the
Majorana LR realization, in which
the Dirac matrix enters the seesaw relation rather than describing light Dirac
neutrinos directly
\cite{Nemevsek:2012iq,Senjanovic:2018xtu,Kiers:2022cyc}. The present work extends the parity-based RH-mixing reconstruction
to the minimal doublet model with purely Dirac neutrinos, where the solution space is qualitatively
different: the lightest neutrino mass becomes a control parameter and the discrete
sign branches determine whether RH leptonic mixing remains aligned or becomes
enhanced.

 The resulting framework therefore defines a predictive Dirac benchmark within LR
symmetry, characterized by a calculable pattern of RH leptonic mixing
that differs qualitatively from both the quark sector and the Majorana realization.

\section{Minimal Dirac Left-Right setup}

We consider the minimal left--right symmetric model with generalized 
parity $\mathcal P$ and doublet-driven LR breaking, in which neutrinos 
are purely Dirac. The gauge group is $SU(2)_L\times SU(2)_R\times U(1)_{B-L}$ with
$g_L=g_R\equiv g$. The scalar sector contains a bidoublet $\Phi(2,2,0)$
generating fermion Dirac masses and LR-breaking doublets 
$\phi_{L,R}$; parity acts as $\Phi\leftrightarrow\Phi^\dagger$ 
and $\phi_L\leftrightarrow\phi_R$, implying Hermitian Yukawa 
couplings $Y_{1,2}^{(f)}=Y_{1,2}^{(f)\dagger}$ for $f=q,\ell$. 
We take the bidoublet vev $\langle\Phi\rangle=v\,\mathrm{diag}(\cos\beta,\sin\beta\,e^{i\alpha})$. In the lepton sector, the Dirac mass matrices are
\begin{equation}
\begin{split}
M_\nu & =v\!\left(Y_1^{(\ell)}\cos\beta+Y_2^{(\ell)}\sin\beta\,e^{-i\alpha}\right),
\\
M_e&=v\!\left(Y_1^{(\ell)}\sin\beta\,e^{i\alpha}+Y_2^{(\ell)}\cos\beta\right),
\end{split}
\end{equation}
with $Y_{1,2}^{(\ell)}$ Hermitian by $\mathcal P$. The departure of these matrices from Hermiticity is governed by the single parity-breaking parameter
\begin{equation}
\epsilon \equiv \sin\alpha\,\tan 2\beta\,.
\label{eq:eps_def}
\end{equation}
Equivalently, parity implies the coupled constraints
\begin{align}
M_\nu-M_\nu^\dagger &= i\epsilon\!\left(e^{i\alpha}\tan\beta\,M_\nu-M_e\right),\\
M_e-M_e^\dagger &= i\epsilon\!\left(M_\nu-e^{-i\alpha}\tan\beta\,M_e\right),
\end{align}
which make explicit that flavor non-Hermiticity is controlled by the single parameter $\epsilon$.

In the minimal bidoublet Dirac realization, achieving
$m_\nu\ll m_e$ requires a non-generic Yukawa and vev structure. Independently of how this hierarchy is arranged, we show that parity alone already implies predictive and previously unrecognized structure in the right-handed leptonic mixing matrix.

The Dirac mass matrices are diagonalized as
$M_{\nu,e}=U_L^{(\nu,e)}\,m_{\nu,e}\,U_R^{(\nu,e)\dagger}$,
with $m_{\nu,e}$ diagonal and positive.
Parity then correlates the unitary rotations through the mismatch matrices
$U_{\nu}\equiv U_L^{(\nu)\dagger}U_R^{(\nu)}$ and $U_{e}\equiv U_L^{(e)\dagger}U_R^{(e)}$,
so that
\begin{equation}
V_L^{(\ell)}U_\nu = U_e V_R^{(\ell)},
\label{eq:VLUVR}
\end{equation}
with observable leptonic mixings
\begin{equation}
V_L^{(\ell)} = U_L^{(e)\dagger}U_L^{(\nu)},\qquad
V_R^{(\ell)} = U_R^{(e)\dagger}U_R^{(\nu)},
\end{equation}
where $V_L^{(\ell)}$ corresponds to the PMNS matrix.
In the parity limit $\epsilon=0$, both mismatch matrices reduce to diagonal sign matrices, $U_e=S_e$ and $U_\nu=S_\nu$, yielding the aligned solution
\begin{equation}
V_R^{(\ell)} = S_e\,V_L^{(\ell)}\,S_\nu.
\end{equation}
Here $S_e=\mathrm{diag}(s_{e_i})$ and
$S_\nu=\mathrm{diag}(s_{\nu_i})$, with
$s_{e_i}=\pm1$ and $s_{\nu_i}=\pm1$.

Away from the parity limit, for small but nonvanishing $\epsilon$ and in the leptonic hierarchy limit $m_\nu\ll m_e$, the charged-lepton rotation remains
well approximated by its parity-limit form, $U_e = S_e$, while the neutrino rotation, $U_\nu$,
receives nontrivial $\epsilon$-dependent corrections. In this regime, the LR consistency conditions
reduce to a single nonlinear equation governing the neutrino sector. Defining $X\equiv U_\nu m_\nu$, one finds
\begin{equation}
X^2 = m_\nu^2 + i\epsilon\,H\,X,
\label{eq:master}
\end{equation}
where
\begin{equation}
H \equiv V_L^{(\ell)\dagger}\hat m_e\,V_L^{(\ell)},
\end{equation}
and $\hat m_e \equiv S_e m_e$ denotes the diagonal matrix of signed
charged-lepton masses, $\hat m_{e_i}=s_{e_i}m_{e_i}$. 
A derivation of Eq.~\eqref{eq:master} from the full coupled system is given in Appendix~\ref{app:master}, following the parity-reconstruction strategy developed
for the quark sector in Refs.~\cite{Senjanovic:2014pva,Senjanovic:2015yea}.

Once Eq.~\eqref{eq:master} is solved, the neutrino rotation follows as
$U_\nu = X\,m_\nu^{-1}$, and therefore
\begin{equation}
V_R^{(\ell)} = S_e\,V_L^{(\ell)}\,U_\nu,
\end{equation}
so the RH leptonic mixing matrix is fully determined by
$(m_\nu,m_e,V_L^{(\ell)},\epsilon)$ up to discrete signs.

Unitarity of $U_\nu$ further implies
$U_\nu m_\nu - m_\nu U_\nu^\dagger = i\epsilon H$,
from which one obtains the analytic unitarity bound
\begin{equation}\label{eq:emax}
|\epsilon|\le\min_{i,j}\frac{m_{\nu_i}+m_{\nu_j}}{|H_{ij}|}\,.
\end{equation}
This bound provides a conservative estimate of the physical domain; the exact boundary is
determined numerically below. For the input used below, the minimum is controlled approximately by
the $(1,1)$ entry in the normal hierarchy (NH), with the $(3,3)$ entry becoming
relevant as the spectrum becomes compressed, and by the $(3,3)$ entry in the
inverted hierarchy (IH), giving
\begin{equation}
|\epsilon|_{\rm max}^{\rm NH}\sim (5\text{--}10)\,\frac{m_{\nu,\rm lightest}}{m_\tau},
\qquad
|\epsilon|_{\rm max}^{\rm IH}\sim 5\,\frac{m_{\nu,\rm lightest}}{m_\tau}.
\label{eq:emax_scaling}
\end{equation}
To understand the structure of the solutions within this domain,
we now derive analytic approximations of Eq.~\eqref{eq:master}.

 \section{Analytic Structure }

 \paragraph{Small-$\epsilon$ regime.}

For $|\epsilon|\ll1$ the master equation~\eqref{eq:master}
admits a perturbative expansion (see also Ref.~\cite{Senjanovic:2014pva}). The series can be constructed recursively order by order; the explicit recursion relations are given in Appendix~A. To first order the RH mixing matrix takes the form
\begin{equation}
(V_R^{(\ell)})_{ij}
=
s_{e_i}(V_L^{(\ell)})_{ik}
\left[
\delta_{kj}
+
i\epsilon
\frac{H_{kj}}{\hat m_{\nu_k}+\hat m_{\nu_j}}
\right]
s_{\nu_j}
+\mathcal O(\epsilon^2),
\label{eq:VR_eps_PRL}
\end{equation}
with $\hat m_{\nu_i}=s_{\nu_i}m_{\nu_i}$ and $s_{\nu_i}=\pm 1$.

The denominators $(\hat m_{\nu_i}+\hat m_{\nu_j})^{-1}$
can become enhanced for specific sign configurations. In particular, in mixed-sign branches the sum can effectively turn into a difference, and the $(12)$ channel becomes the most sensitive one in practice. For $s_{\nu_1}=-s_{\nu_2}$ the leading
solar-angle deviation, $\delta \theta_{12}\equiv \theta_{12}^R-\theta_{12}^L$ (with angles defined in the standard PDG parametrization), scales as
\begin{equation}
\delta\theta_{12}
\simeq
-\epsilon\,\frac{\Im H_{12}}{\hat m_{\nu_1}+\hat m_{\nu_2}}\,,
\label{eq:dth12_eps_PRL}
\end{equation}
revealing the enhancement associated with small neutrino mass splittings. 
 
 The mixed-sign choice identifies the branches that can be enhanced, but it is the
spectral compression of the corresponding neutrino pair that controls whether the
enhancement becomes parametrically large.

Thus the solar angle is the most sensitive to parity breaking, while the remaining
angles receive only subleading corrections.

\paragraph{Quasi-degenerate regime.}

The small-$\epsilon$ expansion can cease to be reliable in the
presence of near-degenerate neutrino masses, where mixed-sign denominators become small, even though the exact solution of
Eq.~\eqref{eq:master} remains well defined. This motivates a complementary expansion tailored to the quasi-degenerate regime. Expanding around a common neutrino mass scale
$m_0\simeq m_{\nu,\text{lightest}}$,
\begin{equation}
m_{\nu_i}^2 = m_0^2 + \Delta m_{\nu_i}^2,
\qquad
\left|\frac{\Delta m_{\nu_i}^2}{m_0^2}\right|\ll1,
\end{equation}
the master equation can be solved perturbatively in
$\Delta m_\nu^2/m_0^2$. The corresponding recursive construction is summarized in Appendix~A. To leading order the RH mixing matrix reads
\begin{equation}
\begin{split}
(V_R^{(\ell)})_{ij}
=&
s_{e_i}\!
\Bigg[
e^{i\phi_i}\,\delta_{ik}
+ 
\frac{\big(V_L^{(\ell)}\,\frac{\Delta m_{\nu}^2}{m_0^2}\,V_L^{(\ell)\dagger}\big)_{ik}}
     {e^{-i\phi_i}+e^{i\phi_k}}
\Bigg]
(V_L^{(\ell)})_{kj} \\
& -\frac{1}{2}s_{e_i}\,e^{i\phi_i}
\left(V_L^{(\ell)}\frac{\Delta m_{\nu}^2}{m_0^2}\right)_{ij}
+\mathcal O\!\left(\frac{\Delta m_\nu^4}{m_0^4}\right).
\end{split}
\label{eq:VR-Dm2}
\end{equation}

The phases are
\begin{equation}
e^{i\phi_i}
=
s_{\nu_i}\sqrt{1-\epsilon^2\frac{\hat m_{e_i}^2}{4m_0^2}}
+i\,\epsilon\,\frac{\hat m_{e_i}}{2m_0},
\qquad
s_{\nu_i}=\pm1.
\end{equation}

This expression makes the quasi-degenerate physical domain transparent:
the phases remain on the unit circle only for
\begin{equation}
|\epsilon|\leq \min_i \frac{2m_0}{|\hat m_{e_i}|}
= \frac{2m_0}{m_\tau}.
\label{eq:emax_qd}
\end{equation}
In the quasi-degenerate regime, the analytic upper range of $\epsilon$ is therefore
set simply by the common neutrino mass scale and the largest charged-lepton mass,
giving the same estimate for NH and IH.

In contrast with the small-$\epsilon$ regime, the dependence on
$\epsilon$ is now effectively resummed into the phases $\phi_i$.
In the limit $\Delta m_\nu^2\to0$, $V_R^{(\ell)}$ differs from $V_L^{(\ell)}$
only by diagonal phases, while the first RH--LH mixing deviations arise from the off-diagonal correction proportional to
$V_L^{(\ell)}\Delta m_\nu^2 V_L^{(\ell)\dagger}$.
For mixed-sign configurations the denominator
$e^{-i\phi_i}+e^{i\phi_k}$ can become small due to phase
anti-alignment, leading to enhanced RH--LH misalignment.

Focusing on the enhanced branch $s_{\nu_1}=-s_{\nu_2}$,
the leading solar-angle deviation extracted from
Eq.~\eqref{eq:VR-Dm2} behaves as
\begin{equation}
\delta\theta_{12}
\simeq
2s_{\nu_1}\,
\frac{c_{23}^L}{c_{13}^L}\,
\frac{
\Im\!\left[
\big(V_L^{(\ell)}\Delta m_{\nu}^2 \,V_L^{(\ell)\dagger}\big)_{12}
\right]
}{
\epsilon\,m_0\,(\hat m_\mu-\hat m_e)
}.
\label{eq:dth12-mixedsign}
\end{equation}
The hatted charged-lepton masses keep the $S_e$-branch dependence explicit.

Equation~\eqref{eq:dth12-mixedsign} gives the corresponding quasi-degenerate
expression for the enhanced mixed-sign neutrino branch. The apparent
$1/\epsilon$ behavior should not be interpreted as a physical singularity:
as $\epsilon\to0$ the quasi-degenerate expansion leaves its domain
of validity and the perturbative small-$\epsilon$ expansion must be used instead.

Together the small-$\epsilon$ and quasi-degenerate limits describe complementary
asymptotic regions of the enhanced branch structure, while the intermediate
nonlinear regime is determined numerically.
 
 \section{Parameter Space and Branch Structure}

\begin{figure}[t]
\centering
\includegraphics[scale=0.43]{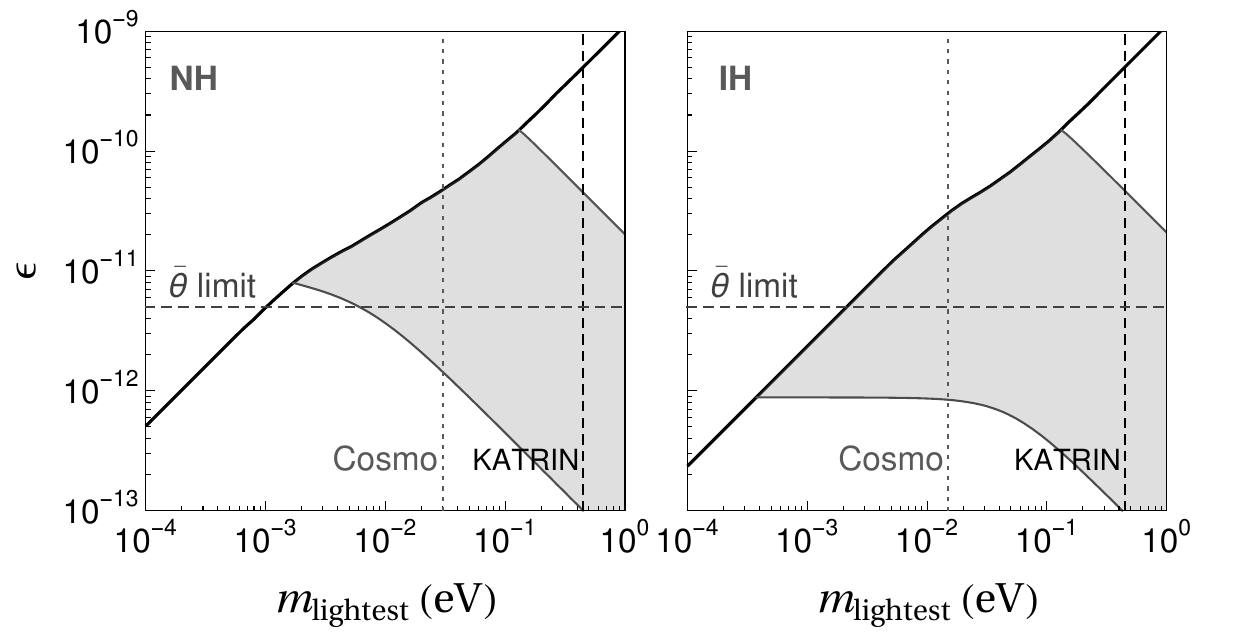}
\caption{
Structure of the $(m_{\nu,\text{lightest}},\,\epsilon)$ plane in the
minimal Dirac LR model for normal (NH, left) and inverted
(IH, right) hierarchies.
The solid black curve shows the maximal $|\epsilon|$ for which
Eq.~\eqref{eq:master} admits physical solutions.
The curves mark the loci where mixed-sign neutrino branches reach
$|\theta_{12}^R-\theta_{12}^L|=8^\circ$.
The shaded band shows the corresponding enhancement region obtained by taking the
union over these mixed-sign branches, where localized RH--LH leptonic mixing
misalignment can become sizable.
Vertical lines indicate the KATRIN and cosmological bounds on the
lightest neutrino mass, while the horizontal line shows the value of
$\epsilon$ implied by the neutron-EDM constraint
$|\bar\theta|\lesssim 10^{-10}$.
Only $\epsilon>0$ is shown, since the full branch set is symmetric under
$\epsilon\to-\epsilon$ once all charged-lepton sign choices are included. }
\label{fig:emax_plots}
\end{figure}

We map the physical domain of the model in the
$(m_{\nu,\text{lightest}},\,\epsilon)$ plane by determining
numerically the maximal value of $\epsilon$ for which
Eq.~\eqref{eq:master} admits physical solutions. In constructing the physical domain and the branch scans, we retain only
solutions for which $U_\nu=Xm_\nu^{-1}$ is unitary and satisfies the original
parity constraints.
The resulting boundary closely follows the conservative
estimate of Eq.~\eqref{eq:emax} and is shown in
Fig.~\ref{fig:emax_plots}. For the numerical analysis we use the current
global-fit values of the neutrino oscillation parameters 
from Ref.~\cite{Esteban:2024eli}.

Within the allowed domain, the phenomenologically
relevant structure is set by the onset of sizable
RH--LH misalignment in the leptonic mixing angles.
As a direct and basis-independent diagnostic, we identify
the locus where
\begin{equation}
|\theta_{12}^R-\theta_{12}^L| = 8^\circ\,.
\label{eq:angle-criterion}
\end{equation}
The numerical value is not a sharp physical threshold, but a representative
choice: it corresponds to a visibly non-aligned RH leptonic mixing pattern, well
separated from the typical quark-like near-alignment outside the enhancement region,
while still low enough to trace the onset of the mixed-sign band in both hierarchies.

For three generations the sign matrices $S_e$ and $S_\nu$ formally give
$2^3\times2^3=64$ assignments. The simultaneous overall sign flip of both matrices
is redundant, leaving 32 distinct $V_R^{(\ell)}$ branches. 
The sign of $\epsilon$ is likewise redundant once the full
charged-lepton sign set is included, so we show only $\epsilon>0$ in Fig.~\ref{fig:emax_plots}.

The $8^\circ$ onset curves are obtained from the analytic
solutions described above. In the small-$\epsilon$
regime we use the perturbative solution of
Eq.~\eqref{eq:VR_eps_PRL}, while in the quasi-degenerate
regime we employ Eq.~\eqref{eq:VR-Dm2}. The two analytic expansions reproduce the numerical solution in their respective
domains of validity and provide a useful description of the boundaries of the
enhancement region.

The resulting band, shown in Fig.~\ref{fig:emax_plots},
originates from mixed-sign configurations in which the
analytic denominators become small.
In the small-$\epsilon$ regime the enhancement is governed
by the neutrino combination
$(\hat m_{\nu_i}+\hat m_{\nu_j})^{-1}$,
whereas in the quasi-degenerate regime the same branch sensitivity is encoded in
the possible phase anti-alignment of
$e^{-i\phi_i}+e^{i\phi_j}$.

The band is obtained by taking the union over mixed-sign neutrino branches
associated with the relevant enhanced pair. The all-equal $S_\nu$ branches remain
smooth and perturbatively aligned and therefore do not contribute to this
enhancement region. The selected branches correspond to genuine
continuations of the parity-limit sign choices.

Outside this band the RH and LH leptonic mixings remain
closely aligned across all branches. Inside it, however,
the mixed-sign neutrino branches can develop localized
but sizable deviations.

The hierarchy of RH--LH angle deviations is strongly non-uniform, with the dominant effect appearing in the solar angle. This behavior can be understood analytically in the two complementary regimes. In the small-$\epsilon$ regime, the enhanced correction is confined to the $(12)$ sector and feeds $\delta\theta_{12}$ already at first order, while the remaining angles are only reached at higher order. In the quasi-degenerate regime, all three angles receive leading corrections, but $\delta\theta_{12}$ still dominates because its numerator receives a contribution from the atmospheric mass splitting, whereas the corrections to $\theta_{13}$ and $\theta_{23}$ are mainly controlled by the solar scale. Thus the dominance of the solar angle reflects a robust structural property of the mixed-sign $(12)$ branches rather than an accidental feature of a particular approximation.

An estimate shows that an enhancement in channel
$(i,j)$ becomes visible only when $|\epsilon|$ exceeds
a characteristic scale of order $|m_i-m_j|/|H_{ij}|$.
The appearance of the effect depends not only
on the mass splittings but also on the mixing
structure encoded in $H_{ij}$. In particular, strongly
hierarchical spectra or small off-diagonal mixings can
push the onset scale beyond the physically allowed $\epsilon$
range.

The global structure of the parameter space can then be understood from the
competition between the onset scale and the maximal allowed value of $\epsilon$.
At low $m_{\nu,\text{lightest}}$, the physical $\epsilon$ range is small and the
relevant neutrino pair may not be sufficiently compressed, so the onset can lie
outside the allowed domain. As the spectrum becomes more compressed, the onset
scale decreases and the enhancement can enter the physical region. This explains
why compressed spectra can display sizable RH--LH misalignment, and why in IH the
effect persists farther into the low-lightest-mass region, since the relevant $(1,2)$
pair is already compressed. For larger $m_{\nu,\text{lightest}}$, the
quasi-degenerate expression \eqref{eq:dth12-mixedsign} describes the large-mass
side of the band. The fixed-angle contour continues toward smaller $\epsilon$,
while in the exact degenerate limit the off-diagonal misalignment vanishes.

This demonstrates that RH leptonic mixing in the minimal
Dirac LR model exhibits a highly structured pattern: significant
misalignment arises only in localized regions controlled
by the interplay of parity breaking and spectral
near-degeneracies, whereas the bulk of parameter space
remains perturbatively stable.

To visualize this structure along a
fixed slice of the parameter space, we examine the behavior
of the RH mixing angles as functions of $\epsilon$. We
extract the angles using the same global-fit values of the neutrino oscillation
parameters.
Since the diagonal external phases are fixed by the master
equation, the resulting angles are unambiguously defined.

\begin{figure}[t]
\centering
\includegraphics[scale=0.43]{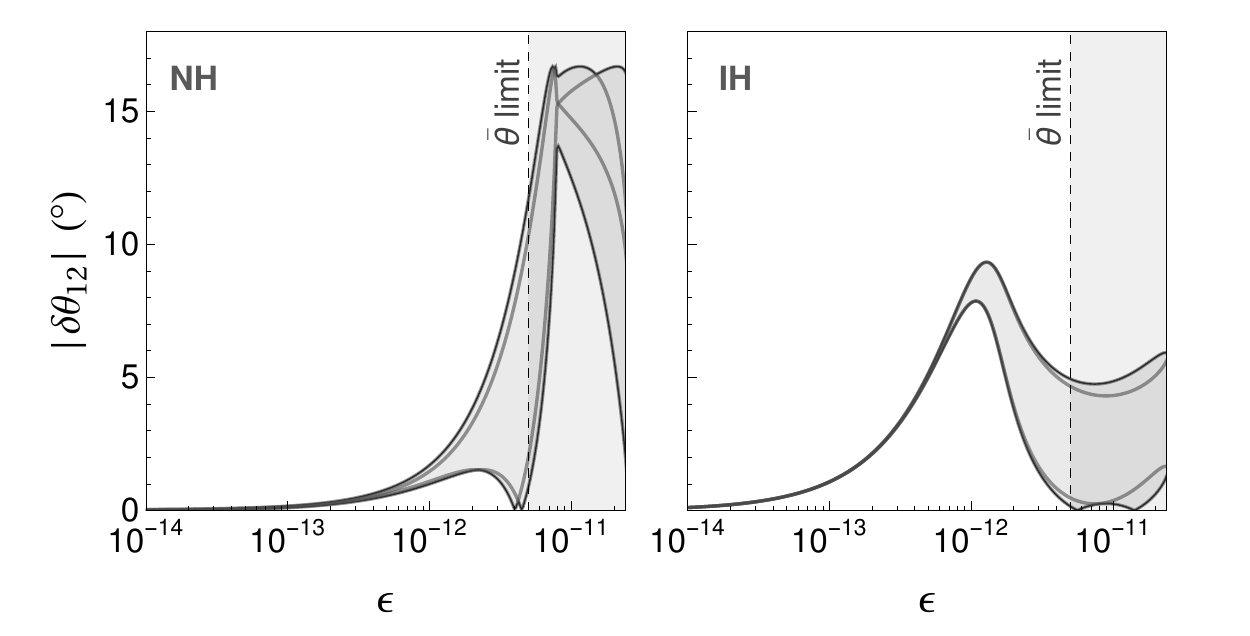}
\caption{
Representative branch behavior of the solar-angle deviation
$|\theta_{12}^R-\theta_{12}^L|$ as a function of positive
$\epsilon=\sin\alpha\tan2\beta$ for normal (NH, left) and inverted
(IH, right) hierarchies at fixed
$m_{\nu,\text{lightest}}=0.01\,\mathrm{eV}$.
Gray curves show the four mixed-sign branches in the $(12)$-enhanced sector ($s_{\nu_1}=-s_{\nu_2}$),
with fixed charged-lepton signs
$S_e=\mathrm{diag}(+,+,-)$, while the black envelope indicates
the extremal RH--LH deviation at fixed $\epsilon$ within
this set of branches.
The vertical dashed line marks the value of $\epsilon$ implied by the neutron-EDM
constraint $|\bar\theta|\lesssim10^{-10}$, and the shaded region lies beyond the
corresponding bound. }
\label{fig:VRangles03}
\end{figure}

Fig.~\ref{fig:VRangles03} shows the branch structure of the
solar-angle deviation for both normal (NH) and inverted (IH)
hierarchies at $m_{\nu,\text{lightest}}=0.01\,\mathrm{eV}$.
The gray curves correspond to the discrete sign branches,
while the black envelope indicates the extremal RH--LH deviation
at fixed $\epsilon$.
The localized peak reflects the mixed-sign enhancement already
identified in the analytic treatment and in the parameter-space map
of Fig.~\ref{fig:emax_plots}.

The behavior differs qualitatively between the two hierarchies.
In the IH case, the enhancement lies entirely within the
physically allowed range of $\epsilon$.
In contrast, for NH it occurs at larger values of $\epsilon$,
so that the $\bar\theta$ constraint intersects the enhancement
region before it fully develops.
As a result, both hierarchies reach comparable maximal deviations,
but the NH profile is effectively truncated by the $\bar\theta$ bound,
while the IH case exhibits the full enhancement structure.

For the benchmark shown in Fig.~\ref{fig:VRangles03}, fixing the
charged-lepton sign choice leaves four distinct mixed-sign branches
in the $(12)$-enhanced sector. Their spread opens up only after the
peak, illustrating that the RH--LH misalignment is not only localized
in $\epsilon$ but also branch dependent.

\section{Phenomenological constraints}

\paragraph{Direct neutrino-mass limits.}
Direct kinematic measurements of tritium $\beta$ decay constrain the effective neutrino mass. The current \textsc{KATRIN} bound $m_\beta<0.45~\mathrm{eV}$ (90\% C.L.)~\cite{KATRIN:2024cdt} implies $m_{\rm lightest}<0.45~\mathrm{eV}$, shown in Fig.~\ref{fig:emax_plots}.

\paragraph{Cosmological limits.}
Cosmological analyses constrain the neutrino mass sum at the $\mathcal{O}(0.1~\mathrm{eV})$ level, e.g.\ $\Sigma\lesssim0.12~\mathrm{eV}$ from CMB+BAO~\cite{Planck:2018vyg}. This translates into a corresponding bound on $m_{\rm lightest}$, shown in Fig.~\ref{fig:emax_plots}. As it depends on cosmological assumptions, we show it for reference.

\paragraph{Implications for strong CP.}

In left--right theories with generalized parity, the strong CP phase
$\bar\theta$ vanishes in the parity-symmetric limit and becomes
calculable after spontaneous parity breaking
\cite{Mohapatra:1978fy,Maiezza:2014ala,Bertolini:2019out,Maiezza:2020fcv}.
At leading order one finds
\begin{equation}
\bar\theta \simeq \epsilon\,\frac{m_t}{2m_b},
\end{equation}
up to subleading corrections \cite{Maiezza:2014ala,Bertolini:2019out}.  The neutron electric dipole moment
constrains $|\bar\theta|\lesssim10^{-10}$ \cite{Pospelov:2005pr,Engel:2013lsa}. 
The resulting bound on $\epsilon$ is indicated in Fig.~\ref{fig:emax_plots}
and in Fig.~\ref{fig:VRangles03}.

In the present Dirac-neutrino framework $\epsilon$ is not free but is bounded by
the requirement that Eq.~\eqref{eq:master} admit physical solutions
\begin{equation}
|\epsilon|\lesssim\epsilon_{\max}(m_{\rm lightest})
\sim \mathcal{O}\!\left(\frac{m_\nu}{m_\ell}\right).
\end{equation}
Thus the leptonic consistency conditions directly translate into a strong 
suppression of the induced strong-CP phase $\bar\theta$, whose maximal size is
controlled by the lightest neutrino mass.

This does not constitute a dynamical solution of the strong CP problem,
since the smallness of $\epsilon$ is not symmetry protected.
Nevertheless, the Dirac doublet framework exhibits a structural
correlation absent in the triplet (Majorana) realization: the
parameter controlling RH leptonic mixing simultaneously suppresses
$\bar\theta$, allowing sizable leptonic misalignment within the
enhancement region while keeping the strong CP phase small.

The situation differs in the triplet (Majorana) realization, where the strong-CP phase can receive additional radiative contributions, leading to further constraints on the leptonic Yukawa sector and on the heavy-neutrino scale \cite{Kuchimanchi:2014ota,Senjanovic:2020int,Li:2024sln}.

\paragraph{Collider implications.} 

The leptonic mixing effects discussed above are challenging to probe directly.
If $W_R$ and the light RH neutrinos were produced, the charged-current decay
$W_R\to \ell_R\nu_R$ would contain the matrix elements
$(V_R^{(\ell)})_{\ell i}$. However, in the purely Dirac realization the RH
neutrinos are effectively massless at collider scales and appear only as missing
energy. For an inclusive measurement at fixed charged-lepton flavor one therefore
sums over the unobserved neutrino species,
\begin{equation}
\sum_i |(V_R^{(\ell)})_{\ell i}|^2 = 1,
\end{equation}
so the mixing angles themselves are not resolved. The enhancement band found above
should therefore be viewed primarily as a structural prediction of the
parity-constrained Dirac leptonic Yukawa sector, rather than as an immediately
accessible collider observable.

Nevertheless, the gauge sector offers a clean collider discriminator of the
left--right symmetry-breaking pattern. In left--right theories the ratio of heavy
neutral and charged gauge-boson masses depends on the scalar representation
responsible for breaking $SU(2)_R\times U(1)_{B-L}$. The quoted numerical ratios assume the parity-symmetric LR limit
$g_R=g_L$, with $g_{B-L}$ fixed by the standard hypercharge matching,
$g_Y^{-2}=g_R^{-2}+g_{B-L}^{-2}$. With this convention, the minimal triplet
realization gives $M_{Z_R}/M_{W_R}\simeq 1.7$, while doublet breaking gives the
smaller ratio $M_{Z_R}/M_{W_R}\simeq 1.2$. This standard representation dependence of the heavy gauge-boson spectrum is used,
for example, in Ref.~\cite{Solera:2023kwt}.

A measurement of the heavy gauge-boson spectrum would therefore directly probe the
nature of left--right symmetry breaking: a ratio near the triplet prediction would
favor the Majorana realization, while a value close to the doublet prediction would
point toward the Dirac scenario studied here. 

Current LHC searches constrain heavy RH gauge bosons to the multi-TeV range through
charged-lepton plus missing-energy and dilepton resonance searches
\cite{ATLAS:2019lsy,CMS:2021ctt}. Representative bounds are
$M_{W_R}\gtrsim 5\,\mathrm{TeV}$ and $M_{Z_R}\gtrsim 4\,\mathrm{TeV}$.
Dijet searches provide complementary constraints on $W_R$ at the few-TeV level
\cite{ATLAS:2019fgd}.
Collider implications and constraints on the gauge and scalar sectors in the
doublet left--right framework have been studied in
Refs.~\cite{Solera:2023kwt,Bernard:2020cyi,Karmakar:2022iip}.

The situation is different in the conventional triplet (Majorana) realization,
where heavy RH neutrinos can make the flavor structure of RH currents accessible through
lepton-number- and lepton-flavor-violating channels, including the
Keung--Senjanovi\'c (KS) same-sign dilepton process \cite{Keung:1983uu} and its use in
probing RH lepton mixing at the LHC \cite{Das:2012ii,Vasquez:2014mxa}. The collider
phenomenology of the KS channel across the heavy-neutrino mass range, including
prompt, displaced and invisible regimes, has been studied in Ref.~\cite{Nemevsek:2018bbt}. Recent full-Run-2 ATLAS and CMS  searches constrain the
corresponding heavy-neutrino realization at the multi-TeV scale, with limits on
$m_{W_R}$ reaching about $5$--$6.4\,\mathrm{TeV}$ depending on the heavy-neutrino
mass and decay channel \cite{CMS:2021dzb,ATLAS:2023cjo}. Scalar-mediated variants involving
the triplet sector provide additional LNV probes
\cite{Maiezza:2015lza}, while related collider probes of the neutrino Dirac
mass matrix have been discussed in trilepton channels
\cite{Helo:2018rll}.

  \section{Conclusions}
  
  Right-handed leptonic mixing in the minimal Dirac left--right model exhibits a branch-dependent enhancement band arising from the near-singular structure of the LR consistency equation. 

Two complementary perturbative regimes control the behavior of the
solutions. For generic spectra and small parity breaking, an expansion
in $\epsilon$ applies and the leading effects arise from denominators
$(\hat m_{\nu_i}+\hat m_{\nu_j})^{-1}$, producing localized
enhancements driven by neutrino mass splittings. In the
quasi-degenerate regime the appropriate expansion is instead organized
around a common neutrino mass scale, where the dependence on $\epsilon$ is
effectively resummed into phase factors and the enhancement is
controlled by phase anti-alignment. 
Together these regimes provide a unified analytic understanding of the enhancement band across the $(m_{\text{lightest}},\epsilon)$ plane. 

A characteristic nonlinear pattern emerges: specific sign branches
develop localized RH--LH mixing enhancements while the bulk of
parameter space remains perturbatively aligned. This behavior reflects the spectral instability of nearly degenerate systems:
small perturbations can induce large rotations when eigenvalues approach each other.
 
The resulting hierarchy of RH--LH angle deviations is strongly non-uniform, with the dominant effect appearing in the solar-angle deviation, making $\theta_{12}$ the most 
sensitive probe of this scenario, whereas $\theta_{13}$ and 
$\theta_{23}$ receive only subleading corrections.

An important structural feature of the Dirac realization is that the
same parameter $\epsilon$ controlling leptonic misalignment also
induces the strong CP phase. The leptonic consistency bound on
$\epsilon$ therefore suppresses $\bar\theta$, bringing it
close to the current experimental limit. At the same time, the gauge 
sector provides a collider discriminator: doublet breaking predicts 
a characteristic heavy gauge boson mass ratio $M_{Z_R}/M_{W_R}\!\approx\!1.2$, 
distinct from the triplet (Majorana) expectation.

The minimal doublet left--right model thus exhibits a tightly
correlated structure linking parity breaking, neutrino spectra,
leptonic mixing, and strong CP. It therefore provides a predictive Dirac
benchmark against which more elaborate left--right constructions can
be systematically compared.

 Future experimental probes of right-handed currents and neutrino properties may therefore provide a direct window into this structure.
 
\section*{Acknowledgment}

 I thank colleagues in Split for useful discussions.

  \appendix

\section{Derivation of the leptonic master equation and analytic recursions}
\label{app:master}

In this Appendix we derive the leptonic master equation and summarize the recursive
constructions underlying the two analytic expansions. For completeness, we briefly
repeat the relevant definitions and notation.

The starting point and perturbative strategy follow the parity-reconstruction method
developed previously for the quark sector
\cite{Senjanovic:2014pva,Senjanovic:2015yea}, while the reduction to the Dirac
leptonic hierarchy, the resulting equation for $X\equiv U_\nu m_\nu$, and the
quasi-degenerate expansion are specific to the present analysis.

\subsection{From parity constraints to the coupled system}
\label{app:coupled}

Starting from the parity constraints in the lepton sector,
\begin{align}
M_\nu-M_\nu^\dagger &= i\epsilon\!\left(e^{i\alpha}\tan\beta\,M_\nu-M_e\right),
\label{eq:app_MnuMd}\\
M_e-M_e^\dagger &= i\epsilon\!\left(M_\nu-e^{-i\alpha}\tan\beta\,M_e\right),
\label{eq:app_MeMd}
\end{align}
we diagonalize the Dirac mass matrices as
\begin{equation}
M_\nu = U_L^{(\nu)}\,m_\nu\,U_R^{(\nu)\dagger},
\qquad
M_e   = U_L^{(e)}\,m_e\,U_R^{(e)\dagger},
\end{equation}
with $m_\nu$ and $m_e$ diagonal and positive. As in the main text, we define the mismatch matrices
\begin{equation}
U_\nu \equiv U_L^{(\nu)\dagger}U_R^{(\nu)},
\qquad
U_e \equiv U_L^{(e)\dagger}U_R^{(e)},
\end{equation}
and the observable leptonic mixings
\begin{equation}
V_L^{(\ell)} = U_L^{(e)\dagger}U_L^{(\nu)},
\qquad
V_R^{(\ell)} = U_R^{(e)\dagger}U_R^{(\nu)}.
\end{equation}
These satisfy
\begin{equation}
V_L^{(\ell)}\,U_\nu = U_e\,V_R^{(\ell)}.
\label{eq:app_Vrelation}
\end{equation}

It is convenient to introduce
\begin{equation}
X_\nu \equiv U_\nu m_\nu,
\qquad
X_e \equiv U_e m_e.
\end{equation}
Using Eq.~\eqref{eq:app_Vrelation} to eliminate $V_R^{(\ell)}$ from
Eqs.~\eqref{eq:app_MnuMd}--\eqref{eq:app_MeMd}, we obtain the coupled system
\begin{align}
X_\nu^2 &=
m_\nu^2
- i \epsilon \!\left(
\tan\beta\,e^{i\alpha} m_\nu^2
- V_L^{(\ell)\dagger} X_e^\dagger V_L^{(\ell)} X_\nu
\right),
\label{eq:app_Xnu_exact}
\\[4pt]
X_e^2 &=
m_e^2
+ i \epsilon \!\left(
\tan\beta\,e^{-i\alpha} m_e^2
- V_L^{(\ell)} X_\nu^\dagger V_L^{(\ell)\dagger} X_e
\right).
\label{eq:app_Xe_exact}
\end{align}
This form is exact and encodes the full leptonic parity constraints.

\subsection{Leptonic hierarchy limit and master equation}
\label{app:master_derivation}

In the leptonic hierarchy regime $m_\nu\ll m_e$, the coupled system simplifies substantially. 
Keeping the nontrivial nonlinear structure intact, one may consistently neglect the purely diagonal
$\mathcal O(\epsilon)$ terms proportional to $\tan\beta\,e^{\pm i\alpha}$ in the rhs of 
Eqs.~\eqref{eq:app_Xnu_exact} and \eqref{eq:app_Xe_exact},
as well as the $\mathcal O(\epsilon\,m_\nu/m_e)$ off-diagonal correction to $X_e$.
For the latter, even using the electron mass gives
\begin{equation}
\frac{m_\nu}{m_e}\lesssim
\frac{0.1\,{\rm eV}}{0.511\,{\rm MeV}}
\simeq 2\times 10^{-7}.
\end{equation}
Since this correction is also proportional to $\epsilon$, it is safely sub-percent
throughout the $\epsilon$ range of interest. These terms only induce tiny diagonal
rephasings and doubly suppressed off-diagonal corrections, and do not affect the
determination of $V_R^{(\ell)}$ at the level of interest, nor can they spoil the
enhancement band discussed in the main text.

To this accuracy one finds
\begin{equation}
X_e = S_e\,m_e \equiv \hat m_e,
\end{equation}
where $S_e$ is a diagonal sign matrix and $\hat m_e$ the corresponding signed charged-lepton mass matrix.

Substituting into Eq.~\eqref{eq:app_Xnu_exact}, one obtains
\begin{equation}
X_\nu^2
=
m_\nu^2
+ i\epsilon\,
V_L^{(\ell)\dagger}\hat m_e V_L^{(\ell)}\,X_\nu.
\end{equation}
Defining
\begin{equation}
H \equiv V_L^{(\ell)\dagger}\hat m_e V_L^{(\ell)},
\end{equation}
and identifying $X_\nu\equiv X$, this yields the master equation quoted in the main text,
\begin{equation}
X^2 = m_\nu^2 + i\epsilon\,H\,X.
\label{eq:app_master}
\end{equation}
Once Eq.~\eqref{eq:app_master} is solved, the neutrino mismatch matrix follows from
\begin{equation}\label{app:eq_Un}
U_\nu = X\,m_\nu^{-1},
\end{equation}
and the RH leptonic mixing matrix is then
\begin{equation} \label{app:eq_VR}
V_R^{(\ell)} = S_e\,V_L^{(\ell)}\,U_\nu.
\end{equation}

\subsection{Recursive expansion for small \texorpdfstring{$\epsilon$}{epsilon}}
\label{app:eps_recursion}

For $|\epsilon|\ll1$, the master equation~\eqref{eq:app_master} can be solved perturbatively by expanding
\begin{equation}
X = \sum_{n=0}^\infty (i\epsilon)^n X^{(n)},
\qquad
X^{(0)} = S_\nu m_\nu \equiv \hat m_\nu,
\end{equation}
with $S_\nu$ a diagonal sign matrix and $\hat m_\nu$ the corresponding signed neutrino mass matrix. Substituting into Eq.~\eqref{eq:app_master} gives the recursion relation
\begin{equation}
X^{(n)}_{ij}
=
\frac{
\big(H\,X^{(n-1)}\big)_{ij}
-
\sum_{k=1}^{n-1}
\big(X^{(k)}  X^{(n-k)}\big)_{ij}
}{
\hat m_{\nu_i}+\hat m_{\nu_j}
},
\quad n\ge1.
\label{eq:app_eps_recursion}
\end{equation}
The RH leptonic mixing matrix then follows from Eq.~\eqref{app:eq_Un} and Eq.~\eqref{app:eq_VR}.

At first order one recovers Eq.~\eqref{eq:VR_eps_PRL}, used in the main text. 
The recursion relation~\eqref{eq:app_eps_recursion} makes explicit that denominators of the form $(\hat m_{\nu_i}+\hat m_{\nu_j})^{-1}$ appear at each order, providing the origin of the branch-dependent enhancement discussed in the main text.

\subsection{Recursive expansion in the quasi-degenerate regime}
\label{app:delta_recursion}

In the quasi-degenerate regime we write
\begin{equation}
m_{\nu_i}^2 = m_0^2 + \Delta m_{\nu_i}^2,
\qquad
\Delta_i \equiv \frac{\Delta m_{\nu_i}^2}{m_0^2},
\qquad
|\Delta_i|\ll1,
\end{equation}
so that
\begin{equation}
m_\nu^2 = m_0^2(\mathbf 1+\Delta),
\qquad
\Delta=\mathrm{diag}(\Delta_1,\Delta_2,\Delta_3).
\end{equation}
It is convenient to parametrize
\begin{equation}\label{app:eq_X}
X
=
m_0\,V_L^{(\ell)\dagger}
\Big(Y^{(0)}+Y^{(1)}+\cdots\Big)
V_L^{(\ell)},
\quad
Y^{(n)}=\mathcal O(\Delta^n).
\end{equation}
Substituting into Eq.~\eqref{eq:app_master} and solving order by order in $\Delta$ gives, at zeroth order,
\begin{equation}
Y^{(0)}_{ij} = e^{i\phi_i}\,\delta_{ij},
\quad
e^{i\phi_i}
=
s_{\nu_i}\sqrt{1-\epsilon^2\frac{\hat m_{e_i}^2}{4m_0^2}}
+i\,\epsilon\,\frac{\hat m_{e_i}}{2m_0},
\label{eq:app_phi}
\end{equation}
where the discrete signs $s_{\nu_i}=\pm1$ label the solution branches.

At first order one finds
\begin{equation}
Y^{(1)}_{ij}
=
\frac{
\big(V_L^{(\ell)}\Delta V_L^{(\ell)\dagger}\big)_{ij}
}{
e^{-i\phi_i}+e^{i\phi_j}
},
\label{eq:app_Y1}
\end{equation}
while higher orders satisfy, for $n\ge2$,
\begin{equation}
Y^{(n)}_{ij}
=
-\frac{
\sum_{k=1}^{n-1}\big(Y^{(k)}Y^{(n-k)}\big)_{ij}
}{
e^{-i\phi_i}+e^{i\phi_j}
}.
\label{eq:app_Yn}
\end{equation}

Using Eq.~\eqref{app:eq_X}, Eq.~\eqref{app:eq_Un} and Eq.~\eqref{app:eq_VR}
one then obtains the quasi-degenerate expansion for the RH leptonic mixing matrix quoted in the main text. To first order in $\Delta m_\nu^2/m_0^2$ this gives Eq.~\eqref{eq:VR-Dm2}.

The two expansions are complementary: the small-$\epsilon$ series provides analytic control at weak parity breaking for generic spectra, while the quasi-degenerate expansion captures the regime where neutrino mass splittings are the perturbative quantities and the $\epsilon$ dependence is effectively resummed into the phases~\eqref{eq:app_phi}.

\bibliographystyle{apsrev4-2}
\bibliography{refs}

\end{document}